\documentstyle[12pt, epsfig]{article}  
\begin{document} 
\baselineskip=21pt

\def\la{\mathrel{\mathpalette\fun <}}
\def\ga{\mathrel{\mathpalette\fun >}}
\def\fun#1#2{\lower3.6pt\vbox{\baselineskip0pt\lineskip.9pt
\ialign{$\mathsurround=0pt#1\hfil##\hfil$\crcr#2\crcr\sim\crcr}}}  

\def\lrang#1{\left\langle#1\right\rangle}

\begin{titlepage}   
\title{ANGULAR STRUCTURE OF ENERGY LOSSES OF HARD JET IN DENSE QCD-MATTER} 
\vskip 1cm 
\author{I.P.Lokhtin and A.M.Snigirev \\ ~ \\ 
{\it Moscow State University, Nuclear Physics Institute, 119899 Moscow, Russia}} 
\date{}
\maketitle 
\vskip 1 cm   
\begin{abstract} 
Angular structure of radiative and collisional energy losses of a hard
parton jet propagating through dense QCD-matter is investigated. 
For small angular jet cone sizes, $\theta_0\la 5^0$,  
the radiative energy loss is shown to dominate over the collisional
energy loss due to final state elastic rescattering of the hard
projectile on thermal particles in the medium. Due to coherent
effects, the radiative energy loss decreases with increasing the
angular size the jet. It becomes comparable with the collisional energy
loss for $\theta_0 \ga 5^0-10^0$.

\bigskip

\noindent 
$PACS$: ~~12.38.Mh, 24.85.+p, 25.75.+r \\ 
$Keywords$:~~quark-gluon plasma, energy losses, gluon radiation, relativistic ion 
collisions \\
\end{abstract}
\end{titlepage}    
\newpage 

\noindent {\large \bf 1. Introduction}
\bigskip   

Jet production, as well as other hard processes, is considered to be
an efficient probe for formation of quark-gluon plasma
(QGP)~\cite{qm96} in future experiments on heavy ion collisions at 
RHIC and LHC~\cite{pqm95}. 
High-$p_T$ parton pair (dijet) from a single hard 
scattering is produced at the initial stage of the collision process 
(typically, at $\la 0.01$ fm/c). 
It then propagates through the QGP formed due to mini-jet production at larger time 
scales ($\sim 0.1$ fm/c) and interacts strongly with comoving constituents in the 
medium. 

The actual problem is to study the energy losses of a hard parton jet 
propagating through dense matter. 
We know two possible mechanisms of energy losses: 
$(1)$ radiative losses due to gluon bremsstrahlung induced by multiple 
scattering~\cite{ryskin,brod,gyul94,baier,mueller,zakharov,BDMSexp,BDMSnew} and 
$(2)$ collisional losses due to the final state interaction 
(elastic rescatterings) of high-$p_T$ partons off 
the medium constituents~\cite{mrow91,thoma91,lok1}. 
Since the jet rescattering intensity strongly increases with temperature,  
formation of a super-dense and hot partonic matter in heavy ion
collisions with initial temperatures up to $T_0 \sim 1$ GeV 
at LHC energies~\cite{eskola94} should result in significantly larger
jet energy losses as compared with the case of hadronic gas at 
$T_h \la 0.2$ GeV or a ``cold'' nuclear matter, 
the other parameters of the medium being kept the same.  

In a search for experimental evidences in favour of the medium-induced 
energy losses a significant dijet quenching (a suppression of high-$p_T$ jet
pairs)~\cite{gyul90} and a monojet/dijet ratio enhancement~\cite{gyulqm95} were 
proposed as possible signals of dense matter formation in ultrarelativistic nuclei
collisions. Suppression of high-$p_T$ particles is also considered as
manifestation of jet quenching in the single-particle spectrum~\cite{gyul92}. 
Another option is to perform a direct jet energy losses measurement in processes where 
a hard parton jet is tagged by ``unquenched'' strongly non-interacting particle like 
$Z$-boson~\cite{kvat95} ($q + g \rightarrow q + Z \rightarrow q + \mu^+ + \mu^-$, 
$q + \overline{q} \rightarrow g + Z \rightarrow g + \mu^+ + \mu^-$) or 
$\gamma$-photon~\cite{wang96} ($q + g \rightarrow q + \gamma$, 
$q + \overline{q} \rightarrow g + \gamma$). 
The possibility that the dilepton mass spectra in the invariant mass
range $1.5 \le M \le 2.5$ GeV/c$^2$ 
are modified due to suppression of correlated semileptonic charm and bottom decays, 
$D\bar{D}~(B\bar{B}) \rightarrow l^+l^-$, 
was recently investigated~\cite{shur97,wang97} as a sign of large energy
losses experienced by $c$ and $b$ quarks~\cite{thoma97}.

All the above phenomena (dijet and monojet production, 
$Z + jet$ and $\gamma + jet$ channels, dilepton yield) 
are important for extracting information about the properties 
of super-dense matter to be created in ultrarelativistic heavy ion
collisions and are related, one way or another, with the medium-induced 
QCD energy losses. 

Although the radiative energy losses of a high energy projectile parton 
have been shown to dominate over the collisional losses by up to an order 
of magnitude~\cite{gyul94,thoma97}, a direct experimental verification 
of this phenomenon remains an open problem. Indeed, with increasing of hard parton
energy the maximum of the angular distribution of bremsstrahlung gluons has shift
towards the parent parton direction. This means~\cite{ryskin} that measuring the jet 
energy as a sum of the energies of final hadrons moving inside an angular cone with a 
given finite size $\theta_0$ will allow the bulk of the gluon radiation to belong to 
the jet and thus the major portion of the initial parton energy to be reconstructed. 
Therefore, the medium-induced radiation will, in the first place, soften 
particle energy distributions inside the jet, increase the
multiplicity of secondary particles, but will not affect the total jet energy. 

Dependence of radiative energy losses on the angular cone size of a jet 
should be studied to allow a meaningful comparison with future
experimental data on jet production in ultrarelativistic ion collisions.  
The aim of the present paper is to analyze the angular
structure of radiative and collisional energy losses of hard parton
jet in a dense QCD matter and to study how the 
radiative and collisional losses affect the total energy of a hard
parton jet of a finite angular cone size. 

It was recently shown~\cite{baier,mueller} that the radiation of
energetic gluons in a QCD medium is essentially different from the
Bethe-Heitler independent radiation pattern. Such gluons have 
formation times exceeding the mean free path for QCD parton scattering in the medium.  
In these circumstances the coherent effects play a crucial r\^ole leading to a strong
suppression of the medium-induced gluon radiation. This coherent suppression is a QCD 
analogue of the Landau-Pomeranchuk-Migdal effect in QED. It is important to notice 
that, contrary to the standard gluon bremsstrahlung in hard processes 
where the radiation angles have a broad logarithmic distribution which 
is practically independent on the gluon energy $\omega$, the coherent LPM 
radiation is concentrated at $\theta\sim \theta(\omega)\ll1$ with
$\theta(\omega)$ a given ($E_{jet}$--independent) function of the gluon energy.  
This induces a strong dependence of the jet energy on the jet cone size $\theta_0$. 

On the other hand, the collisional energy losses represent an incoherent sum over 
all rescatterings. It is also almost independent of the initial parton
energy. At the same time, the angular distribution of the collisional
energy loss is essentially different from that of the radiative one. 
The bulk of ``thermal'' particles knocked out of the dense matter by elastic 
scatterings fly away in almost transverse direction relative to the hard
jet axis. As a result, the collisional energy loss turns out to be
practically independent on $\theta_0$ and emerges outside the narrow jet cone. 

\vskip 1cm 
\noindent {\large \bf 2. Energy losses of a jet with finite cone size: 
  heuristic discussion} 
\bigskip   

\noindent{\it 2.1 Collisional energy losses} 
\bigskip 

In this section we apply simple qualitative considerations in order to analyze 
the angular structure of collisional losses of a hard jet in a dense matter. 
If the mean free path of a hard parton is larger than the screening
radius in the QCD medium, 
$\lambda \gg \mu_D^{-1}$, 
the successive scatterings can be treated as independent.  
The dominant contribution to the differential cross section for
scattering of a parton with energy $E$ off the 
``thermal'' partons with energy (or effective mass) $m_0 \sim 3T \ll E$ 
at temperature $T$ can be written as~\cite{gyul94,thoma97}  
\begin{equation} 
\frac{d\sigma_{ab}}{dt} \cong C_{ab} \frac{2\pi\alpha_s^2(t)}{t^2}, 
\end{equation} 
where $C_{ab} = 9/4, 1, 4/9$ for $gg$, $gq$ and $qq$ scatterings respectively. 
Here $t$ is the transfer momentum squared,    
$\alpha_s$ is the QCD running coupling constant,  
\begin{equation} 
\alpha_s = \frac{12\pi}{(33-2N_f)\ln{(t/\Lambda^2)}} \>,  
\end{equation} 
for $N_f$ active quark flavours, and $\Lambda$ is the QCD scale
parameter which is of the order of the critical temperature,  $\Lambda\simeq T_c$.
The integrated parton scattering cross section, 
\begin{equation} 
\sigma_{ab} = \int\limits_{\displaystyle
\mu^2_D(\tau)}^{\displaystyle m_0(\tau)E / 2 }dt\frac{d\sigma_{ab}}{dt}\>, 
\end{equation} 
is regularized by the Debye screening mass $\mu_D^2$. 
In relativistic kinematics, $E \gg m_0$, 
in the rest system of the target with effective mass $m_0$
we get the following estimate for the transverse $p_T^{t,i}$ and longitudinal
$p_L^{t,i}$ momenta of the incident and ``thermal'' particles: 
$p_T^t \simeq \sqrt{t}$, $p_L^t \simeq t / 2m_0$; 
$p_T^i \simeq -\sqrt{t}$, $p_L^i \simeq E - t / 2m_0$. 
Then the thermal average of the collisional energy loss due to single  
elastic scattering can be estimated as~\footnote{For simplicity we do not consider 
here the possibility of collisional losses due to soft interactions 
of jet partons with the collective QGP modes 
(plasma polarization)~\cite{mrow91,thoma91},
inclusion of which seems to be irrelevant for our semi-qualitative
discourse.} 
\begin{equation}  
\label{nu_col}
\nu = \lrang{\frac{t}{2m_0}} = \frac12\lrang{\frac1{m_0}}\cdot\lrang{t} 
\simeq \frac{1}{4T \sigma_{ab}} 
\int\limits_{\displaystyle\mu^2_D}^
{\displaystyle 3T E / 2}dt\frac{d\sigma_{ab}}{dt}t \>.
\end{equation} 
Scattering angle $\theta_i$ of the incident parton vanishes in the relativistic
limit,  $\tan{\theta_i} =  p_T^i / p_L^i \simeq \sqrt{t} / E
\rightarrow 0$. 
The scattering angle  $\theta_t$ of a struck ``thermal'' particle 
with respect to the initial direction of the fast parton can be
estimated as  
$\tan{\theta_t} = p_T^t /p_L^t \simeq 2 m_0 / \sqrt{t}$.  
The minimal and maximal values of $\tan{\theta_t}$ are 
$\tan{\theta^{max}_t} \simeq 2 m_0 / \mu_D$ and $\tan{\theta^{min}_t} 
\simeq 2 m_0 / \sqrt{0.5 m_0 E}$ respectively. 
It is straightforward to evaluate the average 
$\lrang{\tan{\theta_t}}$ as     
\begin{equation}  
\lrang{\tan{\theta_t}} =  \lrang{\frac{2m_0}{\sqrt{t}}} \simeq \frac{6T}{\sigma_{ab}} 
\int\limits_{\displaystyle \mu^2_D}^
{\displaystyle 3T E / 2} dt \frac{d\sigma_{ab}}{dt} \frac{1}{\sqrt{t}}.  
\end{equation} 
Neglecting a weak $\alpha_s (t)$ dependence we obtain 
$\lrang{\tan{\theta_t}} \simeq \frac{2}{3} \tan{\theta^{max}_t} 
\simeq 4m_0 / 3\mu_D$. 
Substituting  $m_0 = 3T$ and the lowest order pQCD estimate 
$\mu_D^2 \cong 4 \pi \alpha_s^{\ast} T^2 (1 + N_f / 6)$  
we arrive at  $<\theta_t> \sim 60^0$ for $T \ga 200$~MeV. 
This value exceeds typical cone sizes $\theta_0\sim 10^0 - 30^0$ 
used to experimentally define hadronic jets.  
This means that the major part of ``thermal'' particles will fly outside the 
cone of the jet and thus cause the ``jet energy loss''.   
\bigskip 

\noindent{\it 2.2 Radiative energy loss} 
\bigskip 

In \cite{baier,mueller} three regimes of gluon radiation off a 
high energy parton in a QCD matter were discussed:   
the Bethe-Heitler regime of independent emission 
(producing energy loss $dE / dx \propto \mu_D^2$), 
the Landau-Pomeranchuk-Migdal regime 
($dE / dx \propto\sqrt{E}$), 
and the factorization limit 
($dE/dx \propto E$, modulo logarithms). 
The radiation pattern depends on the energy of the radiated gluon 
and on the properties of the medium. 

\paragraph{BH regime.}
Small energy gluons with $\omega\la E_{LPM}=\mu_D^2\lambda_g$ follow the 
Bethe-Heitler regime. Hereafter $\lambda_g$ is the gluon mean free path. 
This radiation produces finite energy loss per unit length, 
is concentrated at large angles,
\begin{equation}\label{BHang}
 \omega\la \mu_D^2\lambda_g\>, \quad k_t\sim \mu_D\>; \qquad
 \theta^{(BH)}\simeq \frac{k_t}{\omega}\ga (\mu_D\lambda_g)^{-1} \equiv
 \theta_M \>\la\>1\>,
\end{equation}
and therefore is similar to the collisional energy loss. 
Formation time of the BH gluons does not exceed the mean free path:
$$
 \tau_f\simeq \frac{\omega}{k_t^2}\>\la\> \lambda_g\>.
$$

\paragraph{LPM regime.}
Radiative energy losses are dominated by emission of relatively energetic
gluons with large formation times $\tau_f\gg\lambda$. 
This is the domain of the coherent LPM regime. 
The structure of the LPM gluon spectrum can be understood,
semi-quantitatively, by equating the gluon formation time with the
time of gluon propagation through the medium in course of which the
gluon transverse momentum accumulates according to the random walk
scattering pattern:
\begin{equation}
  k_t^2\simeq \frac{t}{\lambda_g}\cdot\mu_D^2\>, \quad
  \tau_f\simeq \frac{\omega}{k_t^2}\>=\>t\>,
\end{equation}
where we have taken $\mu_D$ to represent a typical momentum transfer
to the gluon in a single scattering in the medium. 
This gives
\begin{equation}
  \label{eq:tauf}
  \tau_f\simeq \lambda_g \cdot
  \left(\frac{\omega}{E_{LPM}}\right)^\frac12, \qquad 
  k_t\simeq \mu_D\cdot \left(\frac{\omega}{E_{LPM}}\right)^\frac14.
\end{equation}
For the radiation angle we get an estimate
\begin{equation} \label{angen}
  \theta=\theta(\omega) 
  \simeq \theta_M\cdot \left(\frac{E_{LPM}}{\omega}\right)^\frac34.
\end{equation}
Here $\theta_M$ is the characteristic angle depending on the local
properties of the medium, 
\begin{equation}
  \label{eq:thetaMdef}
  \theta_M \>=\> (\mu_D\lambda_g)^{-1}\>,
\end{equation}
the combination that has already appeared in (\ref{BHang}).
Formally, 
\begin{equation}\label{thetaMexp}
\theta_M= \frac{\sigma_g\rho}{\mu_D}\propto \alpha_s^2\cdot 
\left(\rho\mu_D^3\right)\>,  
\end{equation}
where $\sigma_g$ is the gluon scattering cross section and $\rho$ the
density of the scattering centers. 
Strictly speaking, it should be considered a small 
parameter in order to ensure applicability of the independent scattering
picture, see above. 
In reality however $\theta_M$ may be numerically of the order unity
since the momentum scale of the coupling $\alpha_s$ 
in (\ref{thetaMexp}) is small, $\alpha_s=\alpha_s(\mu_D) \la 1$.

\paragraph{Factorization regime.}
Finally, for a medium of a finite size $L$ radiation of the most
energetic gluons is 
{\em  medium-independent}\/ (the factorization component). 
These are the gluons with formation times $\tau_f$ exceeding 
the time $\tau_L=L$ it takes to traverse the medium. 
From (\ref{eq:tauf}) we derive from the condition 
$\tau_f=\tau_L$ the maximal gluon energy in the LPM regime, 
$$
 \omega_{(LPM)} \><\> \left(\frac{L}{\lambda_g}\right)^2 E_{LPM}
\>=\> \frac{\mu_D^2L^2}{\lambda_g}\>.
$$
The gluons that are formed outside the medium carry away a large
fraction of the initial parton energy, proportional to $\alpha_s(E)$.  
This part of gluon radiation produces the standard jet energy profile 
which is identical to that of a jet produced in a hard process in the vacuum. 
Hereafter we shall concentrate on the {\em medium-dependent}\/ effects 
in the angular distribution of energy and will not include the
``vacuum'' part of the jet profile.  

\vskip 1cm 
\noindent {\large \bf 3. A model for energy losses of a jet with finite cone size} 
\bigskip       

We calculate the energy losses for the case which can be realized in symmetric 
ultrarelativistic nucleus-nucleus collisions. Following Bjorken~\cite{bjorken} we treat 
the medium as a boost-invariant longitudinally expanding quark-gluon fluid, and partons 
as being produced on a hyper-surface of equal proper times $\tau = \sqrt{t^2 -  z^2}$. 
Recently the radiative energy losses of a fast parton propagating through expanding 
(according to Bjorken's model) QCD plasma have been derived as $dE / dx | _{expanding} 
= c \cdot dE / dx | _{T_L}$ with numerical factor $c \sim 6$ $(2)$ for a parton created 
outside (inside) the medium, $T_L$ being the temperature at which the dense matter was 
left~\cite{BDMSexp}. We shall denote here this radiative energy losses scenario as 
model $I$. 

We can compare model $I$ with another one which relies on an accumulative energy
losses, when both initial and final state gluon radiation is associated with each 
scattering in expanding medium together including the interference effect by the
modified radiation spectrum as a function of decreasing temperature $dE / dx (T)$
(the model $II$). The total energy loss experienced by a hard parton due to multiple 
scattering in matter is the result of averaging over the dijet production vertex 
($R$, $\varphi$), the momentum transfer $t$ in a single rescattering  
and space-time evolution of the medium:  
\begin{equation}
\Delta E_{tot} = 
\int\limits_0^{2\pi}\frac{d\varphi}{2\pi}\int\limits_0^{R_A}dR\cdot P_A(R)
\int\limits_{\displaystyle\tau_0}^{\displaystyle 
\tau_L}d\tau \left( \frac{dE}{dx}^{rad}(\tau) + \sum_{b}\sigma_{ab}(\tau)\cdot
\rho_b(\tau)\cdot \nu(\tau) \right) .    
\end{equation} 
Here $\tau_0$ and $\tau_L = \sqrt{R_A^2-R^2\sin^2{\varphi}} -
R\cos{\varphi}$ are the proper time of the QGP
formation and the time of jet escaping from the plasma, respectively 
($R_{A}$ is the radius of the nucleus). 
Function $P_A(R) \simeq 3(R_A^2-R^2) / (2 R_A^3)$ at $R \leq R_A$ 
describes the distribution of the distance  $R$ from the nuclear collision
axis $z$ to the dijet production vertex for the uniform nucleon density;     
$\rho_b \propto T^3$ is the density of plasma constituents of type $b$ at 
temperature $T$; $\sigma_{ab}$ is the integral cross section of
scattering of the jet parton $a$ 
off the comoving constituent $b$ (with the same longitudinal rapidity). 

The interval between successive scatterings, 
$l_i = \tau_{i+1} - \tau_i$, is determined in linear kinetic theory 
according to the probability density:  
\begin{equation} 
\frac{dP}{dl_i} = \lambda^{-1}(\tau_{i+1})\cdot 
\exp{(-\int\limits_0^{l_i}\lambda^{-1} (\tau_i + s)ds)},
\end{equation}
where the mean inverse free path is given by
$\lambda_a^{-1}(\tau) = \sum_{b}\sigma_{ab}(\tau) \rho_b(\tau)$. 

The thermal-averaged collisional energy loss $\nu(\tau)$ of a jet 
parton due to single elastic scattering outside the angular cone 
$\theta_0$ is estimated 
according to Eq. (\ref{nu_col}) by imposing an additional restriction 
on the momentum transfer $t < (2 m_0 / \tan{\theta_0)^2}$ in order to
ensure $\theta_t>\theta_0$.  

The energy spectrum of coherent medium-induced gluon radiation 
and the corresponding radiative energy loss, $dE / dx$, 
were analyzed in~\cite{mueller} by means of the
Schr\"odinger-like equation whose ``potential'' is determined 
by the single-scattering cross section of the hard parton in the medium. 
Here we suggest a simple generalization of this result to calculate
the gluon energy deposited outside a given cone $\theta_0$.
To this end we make use of the relation (\ref{angen}) between 
the gluon energy and the characteristic emission angle and mean energy $\bar{\omega}$ 
of radiated gluons: 
\begin{equation} 
\label{eq:omega0}
\bar{\omega} < \omega(\theta_0)\>\equiv\> E_{LPM}
\left(\frac{\theta_M}{\theta_0}\right)^\frac{4}{3}\>, 
\end{equation}
which restriction, according to (\ref{angen}), selects the gluons with 
large radiation angles, $\theta>\theta_0$. 

For the quark jet we obtain~\cite{zakharov,BDMSnew}
\begin{eqnarray} 
\label{radiat} 
\frac{dE}{dx}^{rad} = \frac{2 \alpha_s C_R}{\pi \tau_L}
\int\limits_{\omega_{\min}}^E  
d \omega \left[ 1 - y + \frac{y^2}{2} \right] 
\>\ln{\left| \frac{\sin{(\omega_1\tau_1)}}{\omega_1 \tau_1}\right|} 
\>, \\  
\omega_1 = \sqrt{i \left( 1 - y + \frac{C_R}{3}y^2 \right)   
\bar{\kappa}\ln{\frac{16}{\bar{\kappa}}}}
\quad \mbox{with}\quad 
\bar{\kappa} = \frac{\mu_D^2\lambda_g  }{\omega(1-y)}.
\end{eqnarray} 
Here  
$\tau_1 = \tau_L / (2 \lambda_g)$, and
$y = \omega / E$ is the fraction of the hard parton energy 
carried by the radiated gluon, and 
$C_R = 4/3$ is the quark colour factor.  
A similar expression for the gluon jet can be obtained by substituting 
$C_R=3$ and a proper change of the factor in the square bracket in 
(\ref{radiat}), see~\cite{BDMSnew}.  
The integral (\ref{radiat}) and analogous one for $\bar{\omega}$ are carried out over 
all energies from $\omega_{\min}=E_{LPM}$, the minimal radiated gluon energy in the 
coherent LPM regime, up to initial jet energy $E$. 
 
It is worth noticing that such a treatment is approximate since it
is based on the relation between the gluon radiation angle and energy
which holds only in {\it average}. 
The problem of a rigorous description of the differential angular 
(transverse momentum) distribution of induced radiation is complicated 
by intrinsically quantum-mechanical nature of the phenomenon: large
formation times of the radiation does not allow the direction of the
emitter to be precisely defined~\cite{baier,mueller}. 

We also disregard the final state interaction of the hard parton and the 
secondary gluon in the medium after the gluon has been produced. 
Such interactions, even elastic, could affect the angular distribution  
of the emitted gluon with respect to the parent parton.  
As long as the radiation angles are relatively large,
$\theta>\theta_0\la1$, one would not expect the final state redistribution 
effects to be significant. 
In the case of numerically small jet cone size $\theta_0$ 
an importance of the rescattering effects 
should be analyzed separately. 
However, in the limit of very small $\theta_0$ one would not envisage
large final state effects either. Indeed, with $\theta_0$ decreasing,
the energies of the relevant gluons increase, 
and so do the formation times. 
Gluons with formation times of the order of the size of the medium, 
$\tau_f\sim\tau_L$, have got no spare time left to interact 
with the medium after being radiated. 

In order to simplify calculations (and not to introduce new
parameters) we omit the transverse expansion and viscosity of the
fluid using the well-known scaling Bjorken's solution~\cite{bjorken} 
for temperature and density of QGP at $T > T_c \simeq 200$ MeV: 
$T(\tau) \tau^{1/3} = T_0 \tau_0^{1/3},~~ \rho(\tau) \tau = \rho_0 \tau_0$. 
Let us remark that the transverse flow effect can play an important role in the
formation of the final hadronic state at later stages of
the evolution of a long-living system 
created in an ultrarelativistic nuclei collision. At the same time, 
the influence of the transverse expansion of QGP, 
as well as of the mixed phase at $T = T_c$, 
on the intensity of jet rescattering 
(which is a strongly increasing function of temperature) 
seems to be inessential for high initial temperatures 
$T_0 \gg T_c$~\cite{lok1}. 
On the contrary, the presence of viscosity slows down the cooling rate, 
which leads to a jet parton spending more time in the hottest regions 
of the medium. As a result the rescattering intensity goes up, 
i.e., in fact an effective temperature of the medium 
gets lifted as compared with the perfect QGP case~\cite{lok1}. 
Also for certainty we have used the initial conditions 
for the gluon-dominated plasma formation in central $Pb-Pb$ collisions 
at LHC energies, which have been estimated perturbatively 
in~\cite{eskola94} using the new HERA parton distributions: 
$\tau_0 \simeq 0.1$ fm/c, $T_0 \simeq 1$ GeV, $N_f \simeq 0$~\footnote{These estimates 
are of course rather approximate and model-depending. 
The discount of higher order $\alpha_s$ terms 
in computing the initial energy density of the mini-jet system, 
uncertainties of structure functions in the low-$x$ region, 
and nuclear shadowing  can result in variations of the initial energy density 
of QGP~\cite{eskola94}.}. 

Figure 1 represents the average radiative (coherent medium-dependent part) and 
collisional energy losses of a quark-initiated jet with initial energy $E = 100$ GeV in 
the central rapidity region $y=0$ as a function of the parameter $\theta_{0}$ 
of a jet cone size for $T_0 = 1$ GeV and $R_A = 1.2 \cdot (207)^{1/3} \simeq 7$ fm. 
Let us remark that the choice of the scale for a minimal 
jet energy $E \sim 100$ GeV corresponds to the estimated threshold for 
``true'' QCD-dijet recognition from ``thermal'' background 
--- statistical fluctuations of the transverse energy 
flux in heavy ion collisions at LHC energies~\cite{lok2}. 
We can see the weak $\theta_0$-dependence of collisional losses, 
at least $90 \%$ of scattered ``thermal'' particles 
flow outside a rather wide cone $\theta_0 \sim 10^0 - 20^0$. 
The results for the radiative energy losses in both models $I$ and $II$ are very
similar. The radiative losses are almost independent of the initial jet energy and 
decrease rapidly with increasing the angular size of the jet at $\theta_0 \ga 5^0$.  

\vskip 1cm 
\noindent {\large \bf 4. Conclusions} 
\bigskip 

To summarize, we have considered the angular structure of medium-induced radiative and 
collisional energy losses experienced by a hard parton which is produced before the hot 
dense matter is formed and propagates through an expanding quark-gluon fluid. In our 
analysis we took into account the QCD Landau-Pomeranchuk-Migdal effect and effects of 
the finite volume of the medium. Although the radiative energy loss of a jet with a 
small cone size $\theta_0$ can be much larger than the collisional loss, the former 
decreases considerably with $\theta_0$ increasing. On the other hand, the major part of 
``thermal'' particles knocked out of the dense matter by elastic rescatterings, fly 
outside a typical jet cone of $\theta_0 \sim 10^0 - 20^0$. 

In our model, within the parameter range used, the coherent part of the medium-induced 
radiative loss dominates over the collisional loss by a factor $\sim 3$ at 
$\theta_0 \rightarrow 0$. The radiative loss due to coherent LPM gluon emission is 
comparable with the collisional loss at $\theta_0 \sim 8^0$, and becomes negligible at 
$\theta_0 \ga 10^0$. 

Collisional losses are likely to be significant for jets 
propagating through a hot gluon plasma in the conditions envisaged for central heavy 
ion collisions at LHC energies.     
 
It is a pleasure to thank Yu.L.Dokshitzer for valuable suggestions and comments on this 
work. Discussions with R.Baier, L.I.Sarycheva, D.Schiff and G.M.Zinovjev are gratefully 
acknowledged. 

\begin{figure}[hbtp] 
\begin{center} 
\makebox{\epsfig{file=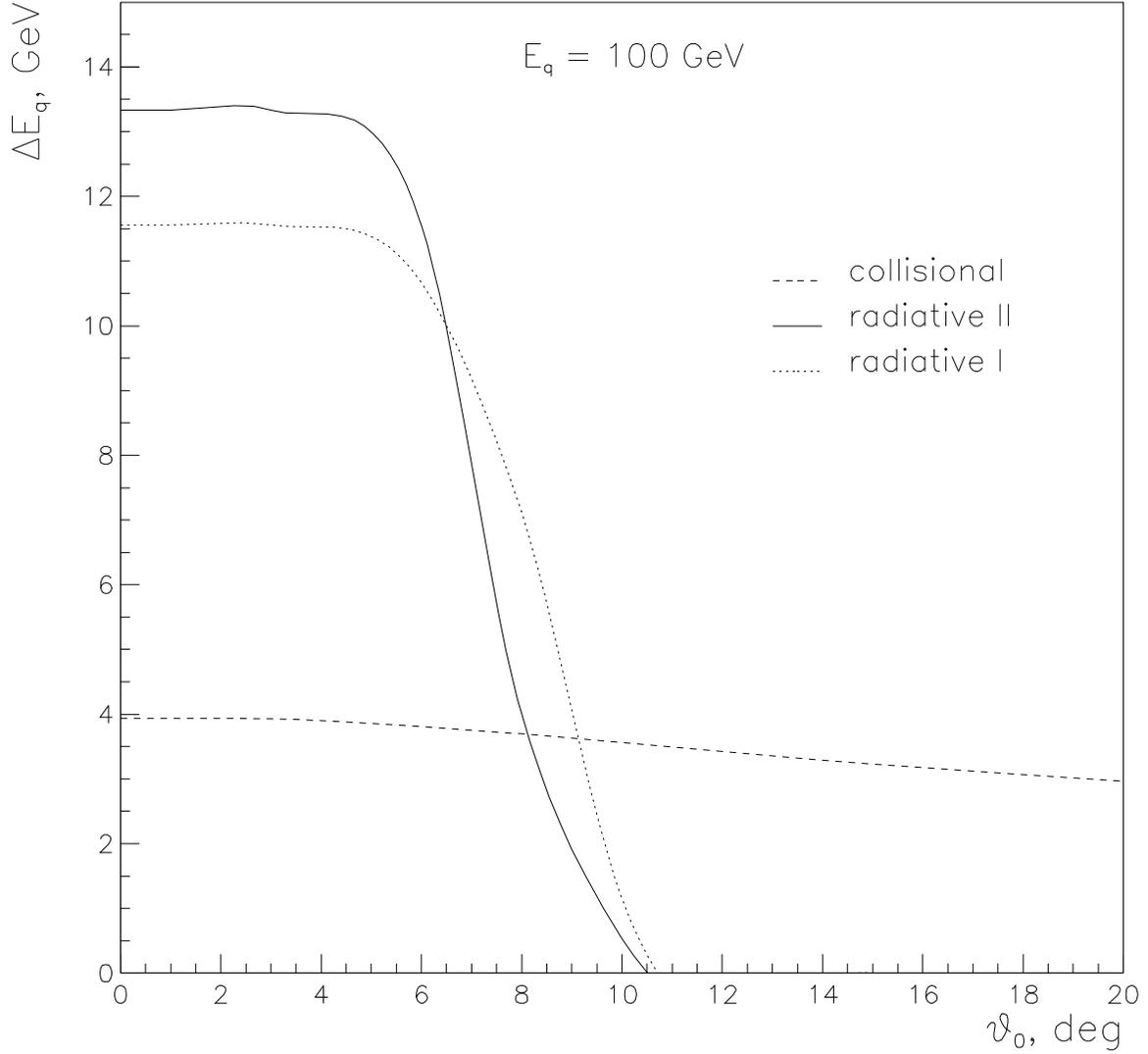, height=170mm}}   
\caption{The average radiative (coherent medium-dependent part, the dotted and solid 
curves for models $I$ and $II$ respectively) and collisional (dashed curves) energy 
losses of quark-initiated jet $\Delta E_{q}$ with initial energy $E = 100$ GeV in the 
central rapidity region $y = 0$ as a function of the parameter $\theta_{0}$ of a jet 
cone size. $R_A = 7$ fm.}  
\end{center}
\end{figure}

\end{document}